\documentclass[prx,twocolumn,showpacs,preprintnumbers,amsmath,amssymb,citeautoscript,superscriptaddress,nobibnotes]{revtex4-1}
\usepackage{amstext}
\usepackage{graphicx}

\makeatletter

\usepackage{dcolumn}
\usepackage{bm}
\usepackage{color}

\makeatother

\begin{document}

\title{Maximizing $\boldsymbol{T_{\rm c}}$ by tuning nematicity and magnetism in FeSe$\boldsymbol{_{1-x}}$S$\boldsymbol{_{x}}$ superconductors}

\author{K.~Matsuura}
\author{Y.~Mizukami}
\author{Y.~Arai}
\author{Y.~Sugimura}
\affiliation{Department of Advanced Materials Science, University of Tokyo, Kashiwa, Chiba 277-8561, Japan}

\author{N.~Maejima}
\author{A.~Machida}
\author{T.~Watanuki}
\affiliation{Synchrotron Radiation Research Center, National Institutes for Quantum and Radiological Science and Technology, Sayo, Hyogo 679-5148, Japan}

\author{T.~Fukuda}
\affiliation{Materials Sciences Research Center, Japan Atomic Energy Agency (SPring-8/JAEA), Sayo, Hyogo 679-5148, Japan}

\author{T.~Yajima}
\author{Z.~Hiroi}
\affiliation{Institute for Solid State Physics, The University of Tokyo, Kashiwa, Chiba 277-8581, Japan}

\author{K.\,Y. Yip}
\author{Y.\,C. Chan}
\author{Q. Niu}
\affiliation{Department of Physics, The Chinese University of Hong Kong, Shatin, Hong Kong}

\author{S.~Hosoi}
\author{K.~Ishida}
\author{K.~Mukasa}
\affiliation{Department of Advanced Materials Science, University of Tokyo, Kashiwa, Chiba 277-8561, Japan}

\author{S.~Kasahara}
\affiliation{Department of Physics, Kyoto University, Sakyo-ku, Kyoto 606-8502, Japan}

\author{J.-G.~Cheng}
\affiliation{Beijing National Laboratory for Condensed Matter Physics and Institute of Physics, Chinese Academy of Sciences, Beijing 100190, China}

\author{S.\,K. Goh}
\affiliation{Department of Physics, The Chinese University of Hong Kong, Shatin, Hong Kong}

\author{Y.~Matsuda}
\affiliation{Department of Physics, Kyoto University, Sakyo-ku, Kyoto 606-8502, Japan}

\author{Y.~Uwatoko}
\affiliation{Institute for Solid State Physics, The University of Tokyo, Kashiwa, Chiba 277-8581, Japan}

\author{T.~Shibauchi}
\email[Email: ]{shibauchi@k.u-tokyo.ac.jp}
\affiliation{Department of Advanced Materials Science, University of Tokyo, Kashiwa, Chiba 277-8561, Japan}

\begin{abstract}
\textbf{A fundamental issue concerning iron-based superconductivity is the roles of electronic nematicity and magnetism in realising high transition temperature ($\boldsymbol{T_{\rm c}}$) \cite{Fernandes14}.  To address this issue,  FeSe is a key material, as it exhibits a unique pressure phase diagram \cite{Sun16} involving nonmagnetic nematic \cite{Baek15} and pressure-induced antiferromagnetic ordered phases  \cite{Bendele10,Terashima15,Kothapalli16,Wang16,Kaluarachchi16,Terashima16}. 
However, as these two phases in FeSe overlap with each other, the effects of two orders on superconductivity remain perplexing. 
Here we construct the three-dimensional electronic phase diagram, temperature ($\boldsymbol{T}$) against pressure ($\boldsymbol{P}$) and isovalent S-substitution ($\boldsymbol{x}$), for  FeSe$\boldsymbol{_{1-x}}$S$\boldsymbol{_{x}}$, in which we achieve a complete separation of nematic and antiferromagnetic phases. In between, an extended nonmagnetic tetragonal phase emerges, where we find a striking enhancement of $\boldsymbol{T_{\rm c}}$. 
The completed phase diagram uncovers two superconducting domes with similarly high $\boldsymbol{T_{\rm c}}$ on both ends of the dome-shaped antiferromagnetic phase. The $\boldsymbol{T_{\rm c}(P,x)}$ variation implies that nematic fluctuations unless accompanying magnetism are not relevant for high-$\boldsymbol{T_{\rm c}}$ superconductivity in this system.}
\end{abstract}

\date{\today}
\maketitle



One of the common aspects among unconventional superconductors, including high-$T_{\rm c}$ cuprates, heavy-fermion, and organic materials, is the appearance of a superconducting dome in the vicinity of magnetic order. This has naturally led to the notion of superconducting pairing mechanism driven by magnetic fluctuations \cite{Moriya00,Monthoux07}.
In iron pnictides, high-$T_{\rm c}$ superconductivity also appears near the antiferromagnetic phase \cite{Fernandes14}, which however is accompanied by the tetragonal-to-orthorhombic structural transition with significant electronic anisotropy (nematicity). This gives rise to new theoretical proposals involving the fluctuations of this electronic nematicity as a glue for the electron pairing \cite{Maier14, Lederer15, Kontani10}. Although enhanced nematic fluctuations of ferro-type ($q=0$) are observed experimentally \cite{Bohmer16}, the antiferromagnetic fluctuations are also enhanced \cite{Shibauchi14}, and thus it is difficult to pinpoint the impact of nematic fluctuations on the superconductivity in iron pnictides.

From this viewpoint, the FeSe-based superconducting system is a suitable material for addressing the importance of nematic fluctuations, as it has a unique phase diagram \cite{Sun16}. At ambient pressure, FeSe shows a nematic transition at $T_{\rm s}=90$\,K without magnetic order down to the lowest temperature \cite{Baek15}. Under pressure, antiferromagnetic order is induced \cite{Bendele10,Kaluarachchi16,Kothapalli16,Wang16,Terashima16}, and the superconducting $T_{\rm c}$ is enhanced by more than a factor of 4 \cite{Sun16,Medvedev10}. Recently, it has been shown that the nematic transition can be tuned to a quantum critical point by isovalent substitution of Se with S, but without inducing magnetic order \cite{Hosoi16}. These results indicate the non-equivalence of physical and chemical pressure in this system. This implies that one can control the magnetism and nematicity independently by these tuning knobs, isovalent substitution and physical pressure, which offers the possibility to disentangle intertwined effects of nematic and magnetic fluctuations on high-$T_{\rm c}$ superconductivity.

\begin{figure*}[t]
	\includegraphics[width=\linewidth]{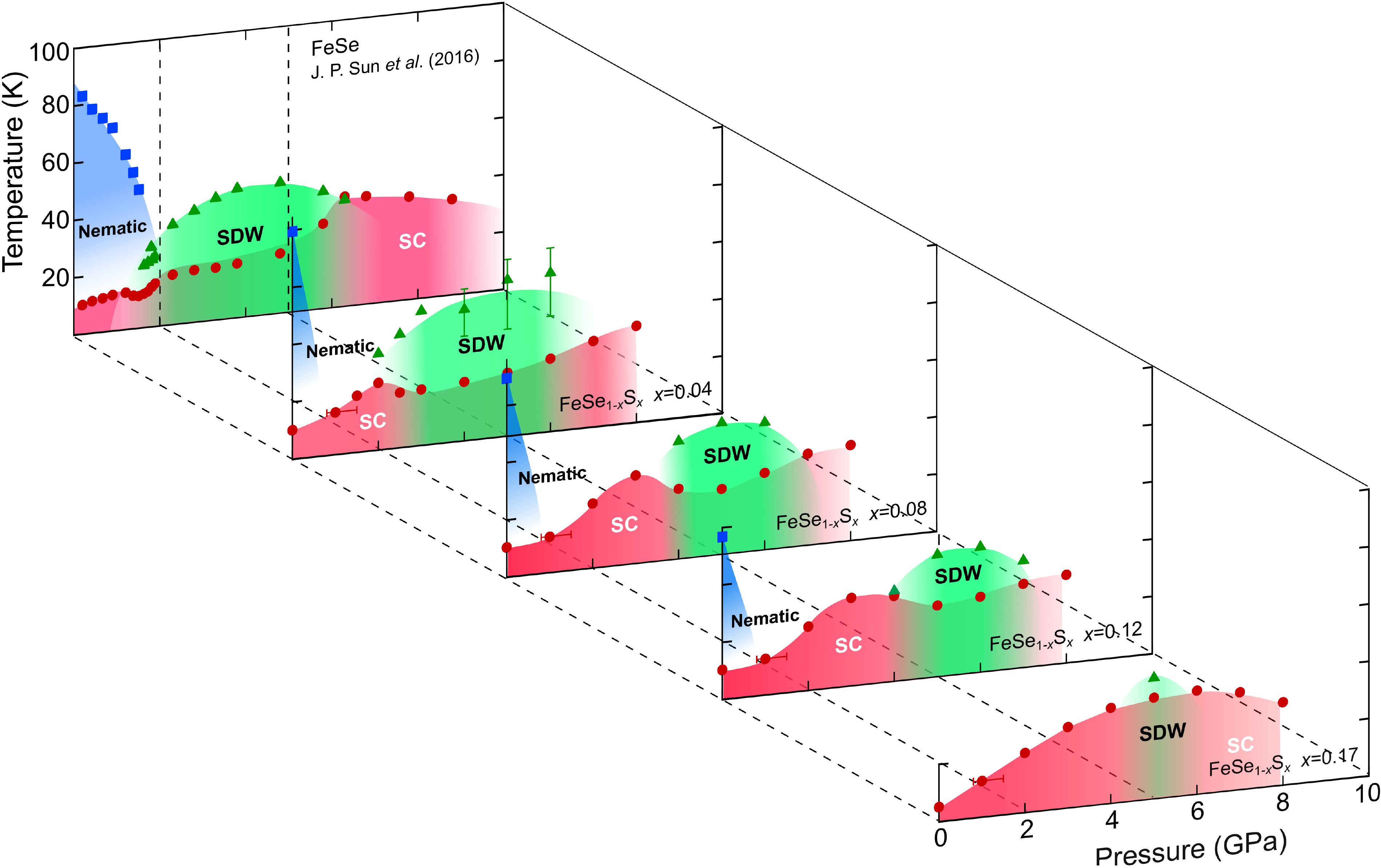}
	\caption{
		\textbf{Temperature-pressure-concentration phase diagram in FeSe$\boldsymbol{_{1-x}}$S$\boldsymbol{_{x}}$.}
		The structural ($T_{\rm s}$, blue squares), magnetic ($T_{\rm m}$, green triangles), and superconducting transition temperatures ($T_{\rm c}$, red circles) are plotted against hydrostatic pressure $P$ and S content $x$. Following the procedure reported for $x=0$ by Sun {\it et al.}\,\cite{Sun16}, $T_{\rm s}$, $T_{\rm m}$, and $T_{\rm c}$ are defined respectively by the temperatures of upturn, kink, and zero resistivity in $\rho(T)$ curves measured in the constant-loading type cubic anvil cell for $x = 0.04$, 0.08, 0.12, and 0.17. The cell is optimized for the high pressure range, and thus for $P<2$\,GPa the error of pressure is relatively large (see error bars for 1\,GPa) compared to higher pressures. The colour shades are the guides for the eyes. Detailed phase diagrams for constant $x$ and $P$ are shown in Figs.\,S1 and S2, respectively. 
	}
	\label{fig1}
\end{figure*}

Here we present our systematic study of temperature-pressure-substitution ($T$-$P$-$x$) phase diagrams of FeSe$_{1-x}$S$_{x}$ in wide ranges of pressure (up to $P\sim8$\,GPa) and sulphur content ($0\le x\le0.17$). The results are summarized in Fig.\,\ref{fig1} (see also Figs.\,S1 and S2 for two-dimensional slices). In pure FeSe, it has been shown by several groups that the nematic transition temperature $T_{\rm s}$ is suppressed by pressure ($P<2$\,GPa) but before the complete suppression of $T_{\rm s}$, antiferromagnetic or spin density wave (SDW) order is induced, resulting in an overlap region of these two phases \cite{Terashima15,Sun16,Kaluarachchi16,Kothapalli16, Wang16}. With increasing $x$, the nematic transition temperature $T_{\rm s}$ is lowered and correspondingly the nematic phase is rapidly suppressed by pressure. However, an opposite trend is found for pressure-induced magnetism: the SDW onset pressure is shifted to higher pressure. These lead to the emergence of the tetragonal nonmagnetic phase in between, which becomes wider with increasing $x$. Most importantly, a new high-$T_{\rm c}$ superconducting dome emerges in the tetragonal phase. Based on the obtained three-dimensional phase diagram, the relative importance of nematic and magnetic fluctuations on superconductivity can be investigated in this system. 



\begin{figure*}[t]
	\includegraphics[width=\linewidth]{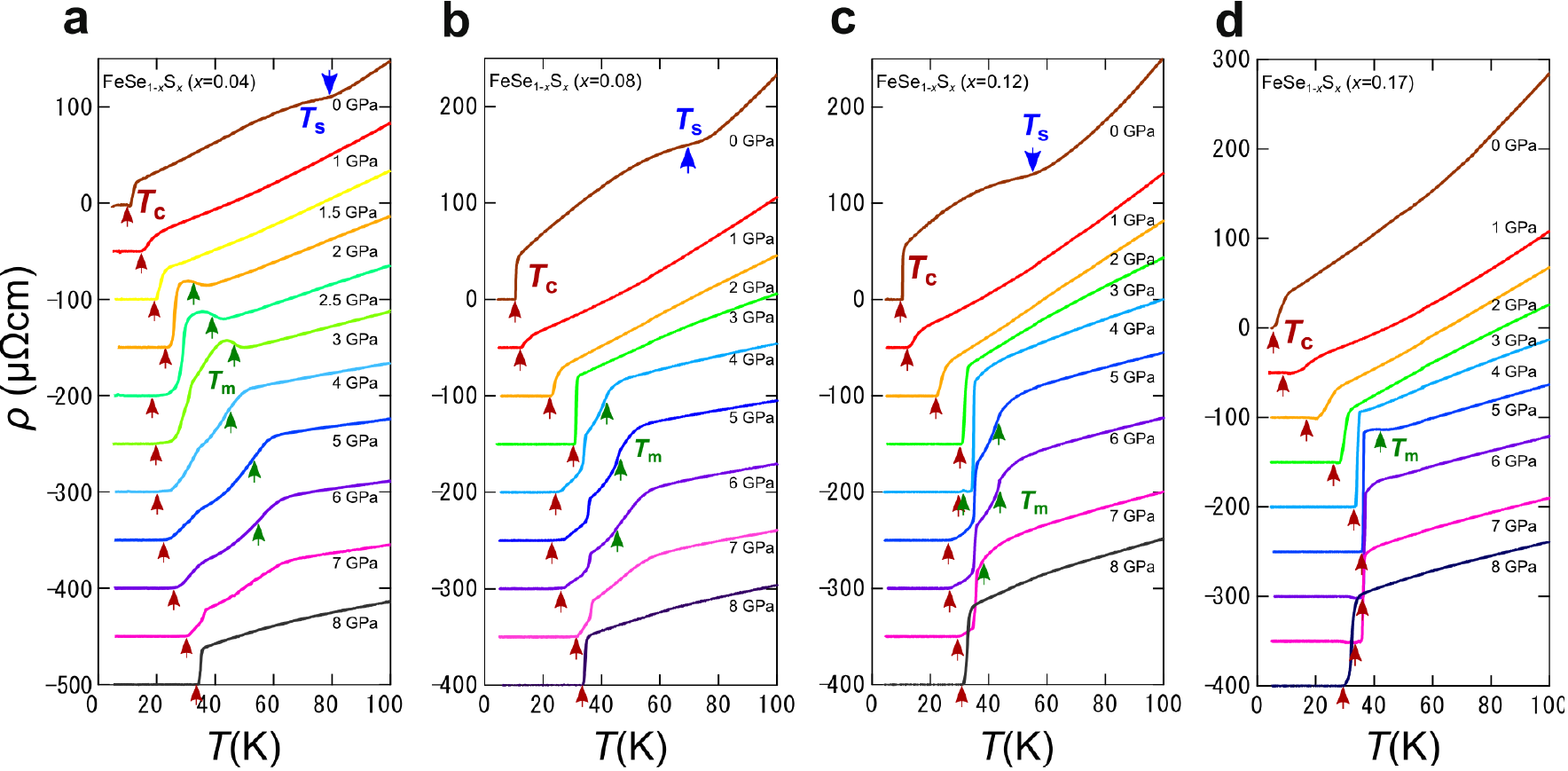}
	\caption{
		\textbf{Evolution of temperature-dependent resistivity under pressure in FeSe$\boldsymbol{_{1-x}}$S$\boldsymbol{_{x}}$.}
		{\bf a-d}, $\rho(T)$ curves below 100 K at different pressures up to 8.0\,GPa measured for $x =0.04$ ({\bf a}), 0.08 ({\bf b}), 0.12 ({\bf c}), and 0.17 ({\bf d}). 
		The data are vertically shifted for clarity. The resistive anomalies at transition temperatures $T_{\rm s}$ (blue), $T_{\rm m}$ (green), and $T_{\rm c}$ (red) are indicated by the arrows. For $x=0.04$ ({\bf a}), the anomalies associated with the magnetic transition is smeared and thus the error of $T_{\rm m}$ determination is relatively large for $P\ge4$\,GPa (see error bars in Fig.\,1). }
	\label{fig2}
\end{figure*}

In FeSe$_{1-x}$S$_x$ at ambient pressure, the temperature dependence of resistivity $\rho(T)$ for $x < 0.17$ exhibits a slight upturn upon cooling at $T_{\rm s}$ due to tetragonal-to-orthorhombic structural transitions, and then it goes to zero below the superconducting transition temperature ($T_{\rm c}$) \cite{Hosoi16}. By measuring $\rho(T)$, we determine the structural transition temperature ($T_{\rm s}$) and $T_{\rm c}$ for $x = 0.04$, 0.08, and 0.12 at ambient pressure as shown in the electronic phase diagram for different S contents (Figs.\,\ref{fig1} and S2a). For $x = 0.17$, we do not observe any signature of the structural transition, indicating the complete suppression of $T_{\rm s}$ as reported previously \cite{Hosoi16}. In Fig.\,\ref{fig2}a-d, we show the evolution of $\rho(T)$ under pressure measured using a cubic anvil cell (CAC) which can generate pressure with a good hydrostatic condition and maintain constant pressure upon cooling \cite{Mori04}. With applying pressure, the $T_s$ anomaly observed at ambient pressure in $x = 0.04$, 0.08, and 0.12 disappears completely at $P\lesssim1$\,GPa. This is a natural consequence of the fact that both S substitution and applying pressure suppress the structural transition in FeSe. 

In $x = 0.04$, the $\rho(T)$ curve at 2.0\,GPa exhibits a clear upturn around 40\,K. The temperature of the upturn increases with pressure, and then it turns to a kink above 4.0\,GPa. This evolution of resistive transition is reminiscent of the magnetic transition seen in FeSe under pressure \cite{Sun16}. Therefore, we follow the procedure of Ref.\,\cite{Sun16} to determine the magnetic transition temperatures ($T_{\rm m}$) by using a dip or peak in d$\rho/$d$T$, and the pressure-evolution of $T_{\rm m}$ is shown in Figs.\,\ref{fig1} and S1b. With increasing pressure, $T_{\rm m}$ is enhanced monotonically up to 6.0\,GPa, while $T_{\rm c}$ is slightly suppressed just after the emergence of magnetism. Above 7.0\,GPa, the kink anomaly due to the magnetic transition disappears, and concomitantly $T_{\rm c}$ determined by the zero resistivity increases gradually up to 32\,K, resembling the evolution of the electronic phases in FeSe at high pressure \cite{Sun16}.

\begin{figure}[t]
	\includegraphics[width=\linewidth]{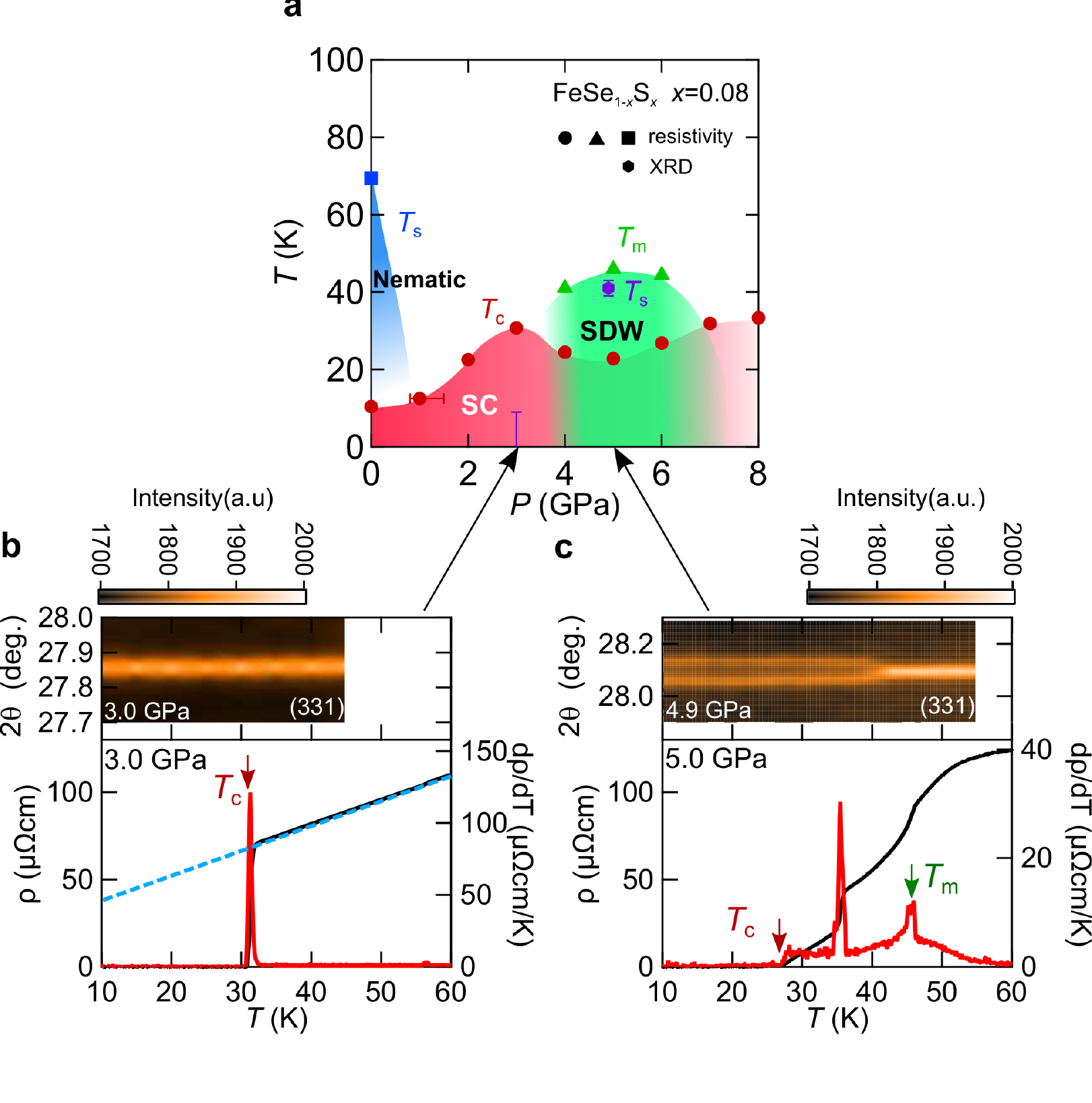}
	\caption{
		\textbf{Temperature-pressure phase diagram for $\boldsymbol{x = 0.08}$. } 
		{\bf a}, $T$-$P$ phase diagram of FeSe$_{1-x}$S$_{x}$ ($x = 0.08$) together with $T_{\rm s}$ determined by the high-pressure synchrotron X-ray diffraction (XRD) in a diamond anvil cell (purple hexagon with error bars). {\bf b},{\bf c}, Temperature dependence of Bragg intensity as a function of $2\theta$ angle is indicated in colour scale for 3.0\,GPa  ({\bf b}) and 4.9\,GPa ({\bf c}). $\rho(T)$ and d$\rho/$d$T$ are also shown with the same horizontal axis. The red (green) arrows indicate $T_{\rm c}$ ($T_{\rm m}$). The blue dashed line in {\bf b} is a $T$-linear fit to the normal-state $\rho(T)$ at 3.0\,GPa. 
	}
	\label{fig3} 
\end{figure}

In $x = 0.08$ and 0.12, we observe remarkable features at moderate pressures. As shown in Fig.\,\ref{fig2}b and Fig.\,\ref{fig2}c, there is no discernible upturn anomaly in $\rho(T)$ between 1.0 and 3.0\,GPa for $x = 0.08$ and 0.12. At 3.0\,GPa, a clear $T$-linear behaviour in the normal-state resistivity is observed (see also Fig.\,\ref{fig3}b),  
which is accompanied by a sharp superconducting transition with enhanced $T_c$ of $\sim32$\,K. We checked for $x=0.12$ that $T_{\rm c}$ determined by ac susceptibility is consistent with that determined by the zero resistivity (see Supplementary Information, Figs.\,S3 and S4). 
Further increase of pressure leads to the emergence of magnetism seen as the kink anomaly around 40\,K, then it persists up to 6.0 (7.0)\,GPa for $x = 0.08$ $(0.12)$. The marked difference compared with FeSe under pressure is the strong enhancement of $T_{\rm c}$ in the lower pressure side of the magnetic phase, forming a peak in $T_{\rm c}$ around 3.0\,GPa for both $x = 0.08$ and 0.12. With increasing pressure above 7.0 (8.0)\,GPa in $x = 0.08$ $(0.12)$, the kink anomaly due to the magnetic transition disappears and $T_{\rm c}$ exhibits another gradual enhancement up to $\sim32$\,K after the disappearance of $T_{\rm m}$, resulting in the double-dome structure in $T_{\rm c}$ having two maxima with almost identical magnitudes. 

In $x = 0.17$, where there is no $T_{\rm s}$ at ambient pressure as shown in Fig.\,\ref{fig2}d, its initial $T_{\rm c}$ of $\sim4$\,K gradually increases up to $\sim35$\,K with pressure, and turns to decrease above 6.0\,GPa, forming a broad superconducting dome as a function of pressure as illustrated in Fig.\,\ref{fig1}. We observe $T_{\rm m}$ only at 5.0\,GPa for this S content, implying that the system is approaching the verge of the pressure-induced SDW phase (Fig.\,S2f).

\begin{figure}[t]
	\includegraphics[width=\linewidth]{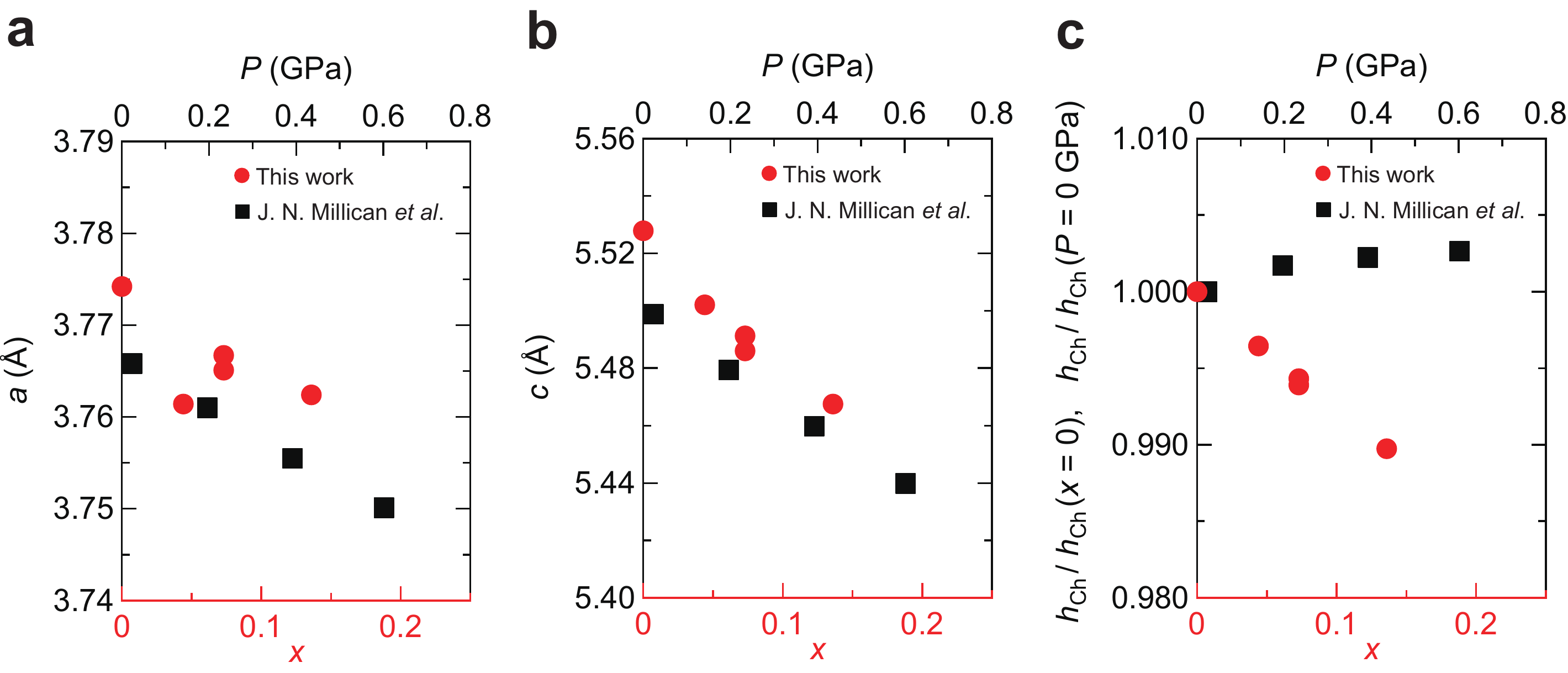}
	\caption{
		\textbf{Comparisons between physical pressure and isovalent substitution effects on the structural parameters.} 
		{\bf a},{\bf b}, Lattice constants $a$ ({\bf a}) and $c$ ({\bf b}) as a function of S content $x$ in the present single crystals of FeSe$_{1-x}$S$_{x}$ (red circles, bottom axis), compared with those as a function of pressure reported for polycrystals of FeSe in Ref.\,\cite{Millican09} (black squares, top axis).  {\bf c}, Chalcogen height $h_{\rm Ch}$ normalised by the initial values as a function of $x$ (red circles, bottom axis) and pressure (black squares, top axis) \cite{Millican09}. The numerical values of these parameters are listed in Table\,SI. 
	}
	\label{fig4} 
\end{figure}

As $x$ is increased, the pressure-induced SDW dome shifts to higher pressure and shrinks, while  low-pressure nonmagnetic phase shifts to lower pressure and disappears at  $x\sim0.17$.  We stress  that the nematic  phase is completely separated from the SDW phase at $x \geq 0.04$.   To confirm the separation between two distinct phases under pressure,  we performed  synchrotron X-ray diffraction measurements under pressure for $x = 0.08$ (Fig.\,\ref{fig3}a).  In Fig.\,\ref{fig3}b,c we show $(331)$ Bragg intensity as a function of temperature at  3.0 and 4.9\,GPa together with the $\rho(T)$ and d$\rho/$d$T$ data. At 3.0\,GPa,  no discernible  change of the Bragg-peak is observed down to the lowest temperature of 10\,K (Figs.\,\ref{fig3}b and S5a). At 4.9\,GPa, on the other hand, the splitting of the Bragg peak is clearly resolved around $T_{\rm s}\sim 41$\,K, evidencing the presence of  the tetragonal-to-orthorhombic structural transition (Figs.\,\ref{fig3}c and S5b).  This structural transition is located very close to the SDW transition at $T_{\rm m}$ at 5.0\,GPa as indicated by the sharp peak in d$\rho/$d$T$ curve in Fig.\,\ref{fig3}b. Thus it is natural to consider that the magnetic phase has an orthorhombic structure, similar to the case of the high-pressure SDW phase of FeSe \cite{Kothapalli16,Wang16}. These results demonstrate that the high-$T_c$ superconductivity in FeSe$_{1-x}$S$_x$ is realised in the tetragonal phase newly emerged between the orthorhombic nematic and magnetic phases. In the nonmagnetic tetragonal phase ($1\lesssim P \lesssim 3$), $T_{\rm c}$ shows a strong increase with $P$ (Figs.\,\ref{fig1} and \ref{fig3}a), indicating that the enhancement of superconductivity is most pronounced near the verge of the magnetic phase, not the nematic phase. It is also likely that the $T$-linear resistivity observed near the SDW boundary (Fig.\,\ref{fig3}b) is a consequence of enhanced antiferromagnetic fluctuations, as reported in other high-$T_{\rm c}$ cases \cite{Shibauchi14,Kasahara10,Sun17}. 

Why the effects of two tuning parameters, physical pressure and isovalent substitution, are so different? In general, applying pressure reduces lattice constants, and it leads to an increase of bandwidth as well as a change in the Coulomb interactions \cite{Scherer17}, often affecting the ground state of the system. The chemical substitution by smaller ions also leads to a decrease of lattice constants, which results in similar effect on the system as the pressure effect. Indeed, in BaFe$_2$As$_2$ system, the physical and chemical pressure effects on superconductivity are essentially similar \cite{Klintberg10}. To address the origin of the difference between chemical and physical pressure effect in FeSe, we determine the structure parameters of FeSe$_{1-x}$S$_x$ at room temperature by single-crystal X-ray diffraction, which are compared with the published data under pressure \cite{Millican09} (Fig.\,\ref{fig4}a-c and Table\,SI). As expected, both $a$- and $c$-axis lattice constants decrease with S-content $x$, which follow the trends under physical pressure. The quantitative comparison suggests that 10\% substitution corresponds to $\sim 0.3$\,GPa (Fig.\,\ref{fig4}a,b). This can be compared with effects of chemical and physical pressure on the phase diagrams of BaFe$_2$As$_2$, where the 30\% substitution of P for As \cite{Kasahara10} and application of $\sim0.55$\,GPa \cite{Colombier09} both lead to the maximum $T_{\rm c}\sim 30$\,K. In sharp contrast to the $a$- and $c$-axis lattice constants, there is a significant difference in the trends of the chalcogen height $h_{\rm Ch}$ from the iron plane (Fig.\,\ref{fig4}c). It has been pointed out in iron pnictides that the height of the anion atoms from the Fe plane plays an important role on the existence of hole-like Fermi surface around the zone corner of the unfolded Brillouin zone, which has significant influence on the nesting properties between the Fermi surfaces \cite{Kuroki09}. It has also been shown that the chalcogen height in FeSe$_{1-x}$Te$_x$ is an important factor for the magnetic interactions \cite{Moon10}. We find that the isovalent substitution reduces $h_{\rm Ch}$ monotonically which is opposite to the observed increasing trend due to physical pressure effect. Thus, it is likely that this difference is responsible for the absence or presence of the induced SDW phase. Indeed, recent theoretical calculation investigating the pressure effect in FeSe points out that the increase of $h_{\rm Ch}$ results in the appearance of Fermi surface in the Brillouin zone corner, which explains the emergence of magnetism under pressure \cite{Yamakawa17}. 

The most notable feature is that the high-$T_c$ superconductivity in the tetragonal phase emerges at the verge of both side of the SDW dome, while $T_c$ is little influenced by the nonmagnetic nematic phase. These results lead us to consider that the ferro-type ($q=0$) nonmagnetic nematic fluctuations do not act as a pairing glue that help increase $T_{\rm c}$, in stark contrast to the nematicity accompanying antiferromagnetic fluctuations ($q\ne0$). In view of the orthorhombicity found in the pressure-induced SDW phase, an intriguing issue that deserves further studies is whether the nematic and magnetic fluctuations cooperatively promote the superconducting pairing, as recently suggested theoretically as ``orbital$+$spin composite fluctuations'' \cite{Yamakawa17}.

\section*{Methods}

High-quality  single crystals of FeSe$_{1-x}$S$_{x}$ ($x = 0$, 0.04, 0.08, 0.12, and 0.17) have been grown by the chemical vapour transport technique \cite{Hosoi16}. The $x$ values are determined by the energy dispersive X-ray spectroscopy. In the crystals obtained under identical conditions, quantum oscillations have been observed in a wide range of $x (\le0.19)$ \cite{Coldea17}, indicating superior crystal quality.  
High-pressure resistivity $\rho(T,P)$ measurements have been performed under hydrostatic pressures up to 8\,GPa with a constant-loading type cubic anvil apparatus which can maintain a nearly constant pressure over the whole temperature range from 300\,K to 2\,K \cite{Sun16,Mori04}. For all these high-pressure resistivity measurements, we employed glycerol as the pressure-transmitting medium, and used the conventional four-terminal method with current applied within the $ab$ plane.
High-pressure ac susceptibility measurements have been done by using a mutual inductance technique in a moissanite anvil cell with glycerol as the pressure-transmitting medium \cite{Klintberg10}. The pressure achieved was determined by measuring the wavelength of the $R_1$ peak of ruby fluorescence.
Synchrotron X-ray diffraction measurements under pressure have been performed at BL22XU in SPring-8 by using diamond anvil cell diffractometer equipped with a gas membrane for maintaining constant pressure on cooling \cite{Watanuki07}. Helium is used as the pressure-transmitting medium. The pressure value in the sample space is monitored by tracking the ruby fluorescence wavelength for the whole temperature range. 

\section*{Acknowledgements}
We thank H. Kontani, and Y. Yamakawa for fruitful discussions. We also thank S. Nagasaki for technical assistance. This work was performed using facilities of the Institute for Solid State Physics, the University of Tokyo. A part of this work was performed under the Shared Use Program of JAEA and QST Facilities (Proposal No. 2015A-E16, 2016A-E16, and 2016B-H13) supported by JAEA, QST Advanced Characterization Nanotechnology Platform as a program of "Nanotechnology Platform" of the Ministry of Education, Culture, Sports, Science and Technology (MEXT), Japan (Proposal No. A-15-AE-0016, A-16-QS-0008, and A-16-QS-0025). The synchrotron radiation experiments were performed by using a QST experimental station at JAEA beamline BL22XU in SPring-8 with the approval of the Japan Synchrotron Radiation Research Institute (JASRI) (Proposal No. 2015A3701, 2015A3783, 2015B3701, 2016A3751, 2016A3781, and 2016B3785). This work was supported by Grant-in-Aids for Scientific Research (A), (B), (S), (Proposal No. 15H02106, 	15H03681, 15H03688, and 25220710), Grant-in-Aids on Innovative Areas ``Topological Materials Science'' (No. 15H05852), CUHK Startup Grant (No. 4930048), and Research Grant Council of Hong Kong (ECS/24300214, GRF/14301316). J.-G. C. acknowledges the supported from the NSFC, MOST, and CAS (Proposal No. 11574377, 2014CB921500, XDB07020100, and  QYZDB-SSW-SLH013).

\section*{Author contributions} 

T.S. conceived the project. K.Matsuura, Y.A., Y.S, J.-G.C. and Y.U. measured the resistivity under pressure using CAC. K.Matsuura, Y.Mizukami, N.M., A.M., T.Watanuki, T.F., S.H., K.I., and S.K. performed high-pressure X-ray diffraction measurements. Y.Mizukami, T.Y., and Z.H. performed X-ray diffraction measurements for FeSe$_{1-x}$S$_x$ at ambient pressure. K.Y.Y., Y.C.C., Q.N., and S.K.G. measured the susceptibility under pressure. K.Matsuura, K.Mukasa, T.Watashige, S.K., and Y.Matsuda synthesized FeSe$_{1-x}$S$_x$ single crystals. All authors discussed the results. K.Matsuura, Y.Mizukami, Y.Matsuda, T.S. wrote the paper with inputs from all authors. Y.Matsuda, Y.U., and T.S. supervised the projects.

\vfill
\renewcommand{\figurename}{Figure S$\!\!$}
\renewcommand{\tablename}{Table S$\!\!$}
\setcounter{figure}{0}

\section*{Supplementary Information}

\subsection{Phase diagrams}

Here we show the three-dimensional temperature-pressure-substitution ($T$-$P$-$x$) phase diagram of Fig.\,1 in two sets of two-dimensional slices; constant-$x$ (Fig.\,S1) and constant-$P$ phase diagrams (Fig.\,S2). In both cases, it is clearly seen that the superconductivity is suppressed inside the SDW phase and the suppression of magnetism is accompanied by the enhancement of $T_{\rm c}$. This evidences the competition between the magnetic order and superconductivity, whereas the fluctuations outside the SDW order help to enhance superconductivity.  In contrast, near the verge of the nonmagnetic nematic phase, $T_{\rm c}$ is quite low.

\begin{figure}[b]
	\vspace{1cm}
	\includegraphics[width=\linewidth]{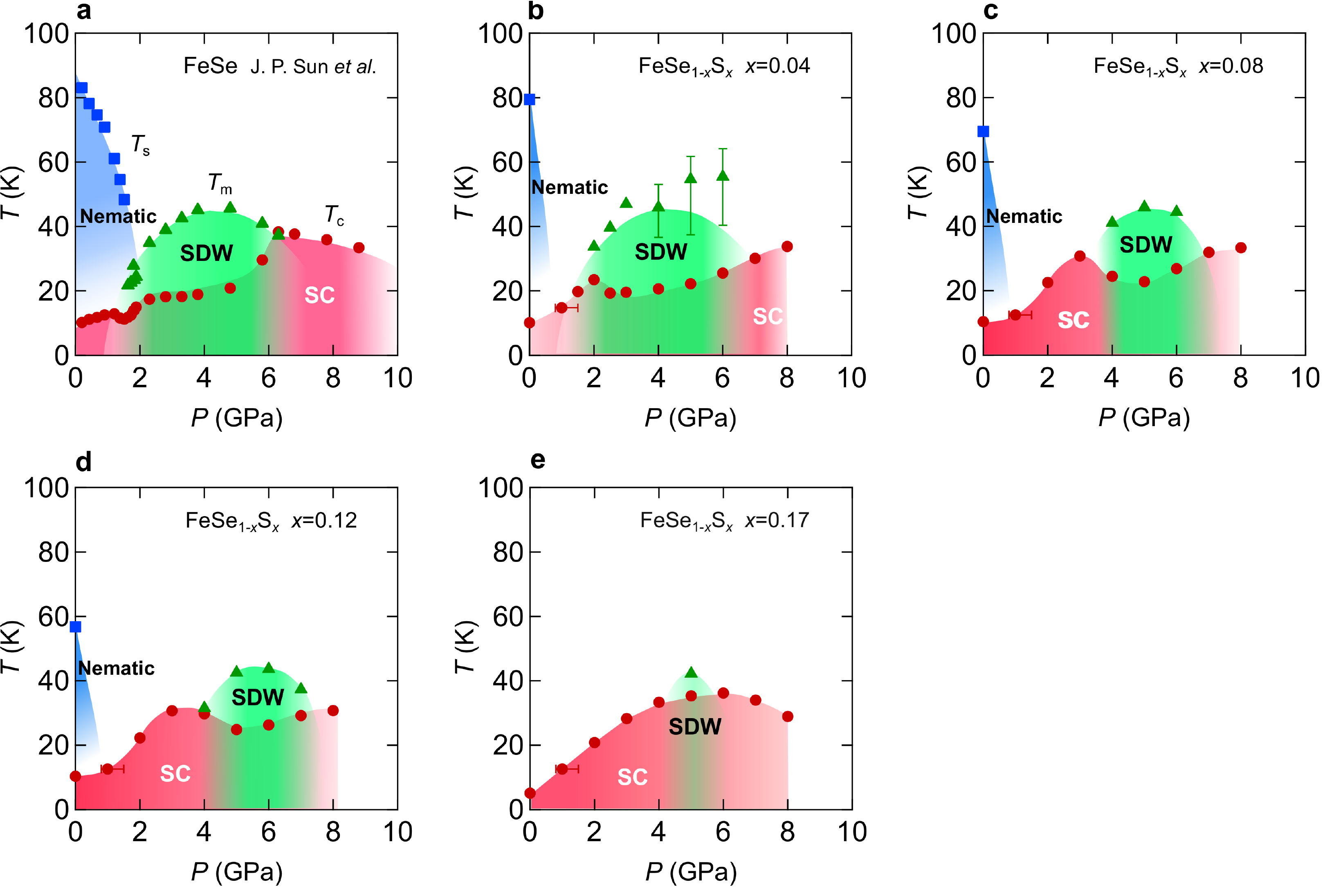}
	\caption{
		\textbf{Temperature versus pressure phase diagrams for constant $\boldsymbol{x}$ values in FeSe$\boldsymbol{_{1-x}}$S$\boldsymbol{_{x}}$.} {\bf a-e}, Pressure dependence of nematic (blue squares), SDW (green triangles), and superconducting transition temperatures (red circles) at $x=0$ ({\bf a}) \cite{Sun16}, 0.04  ({\bf b}), 0.08 ({\bf c}), 0.12 ({\bf d}), and 0.17 ({\bf e}). The colour shades are the guides for the eyes.
	}
\end{figure}

\begin{figure}[t]
	\includegraphics[width=\linewidth]{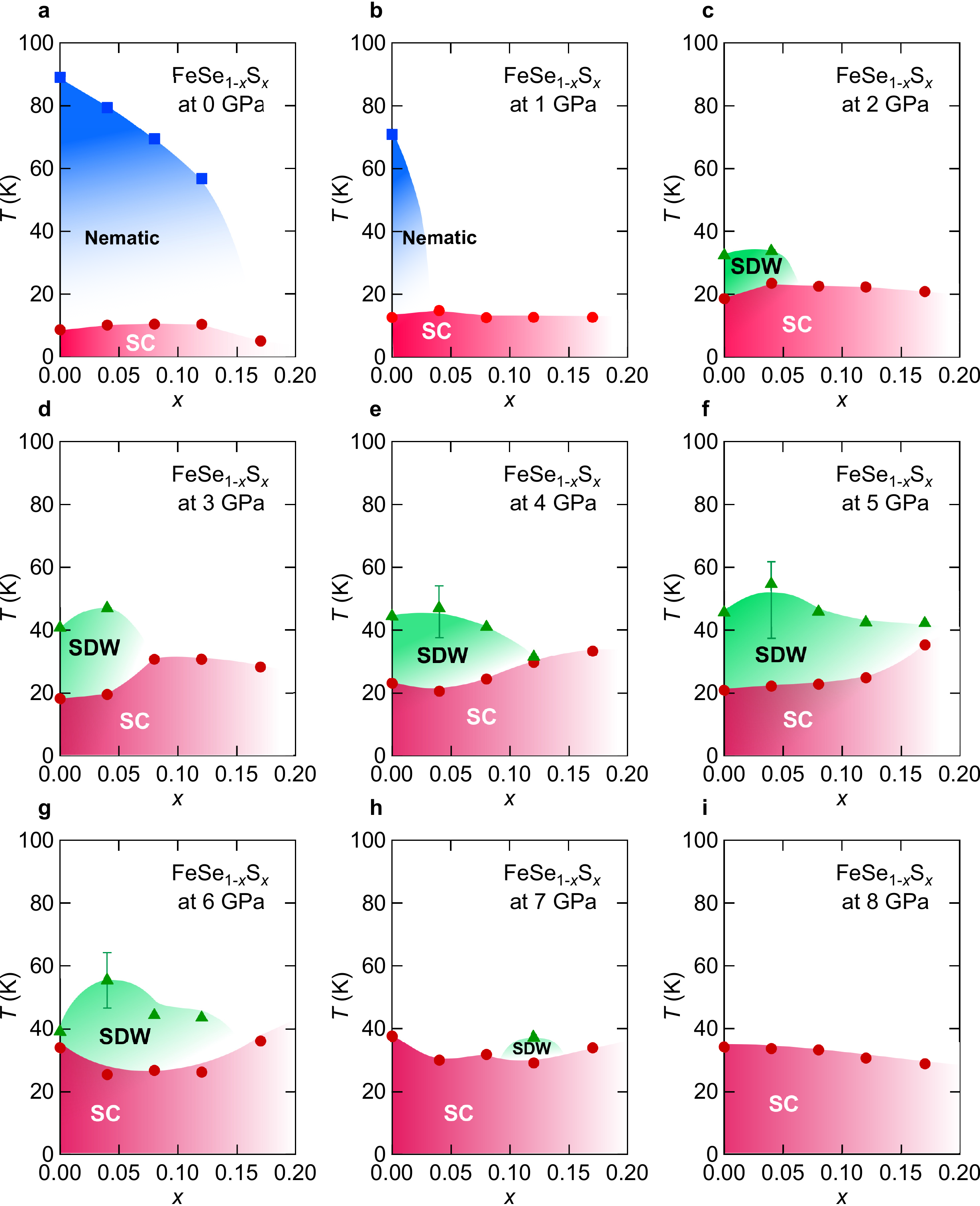}
	\caption{
		\textbf{Temperature versus S-content phase diagrams for constant pressure values in FeSe$\boldsymbol{_{1-x}}$S$\boldsymbol{_{x}}$.} {\bf a-i}, $x$-dependence of nematic (blue squares), SDW  (green triangles), and superconducting transition temperatures (red circles) at $P=0$ ({\bf a}), 1.0 ({\bf b}), 2.0 ({\bf c}), 3.0 ({\bf d}), 4.0 ({\bf e}), 5.0 ({\bf f}), 6.0 ({\bf g}), 7.0 ({\bf h}), and 8.0\,GPa ({\bf i}). The colour shades are the guides for the eyes.
	}
\end{figure}

\newpage 

\subsection{ac susceptibility measurements under pressure}

The superconducting transition at high pressure is also checked by the ac susceptibility measurements for $x=0.12$. The sample is placed in a microcoil inside a self-clamped moissanite anvil cell with glycerol as the pressure medium, and the temperature dependence of the real part of ac susceptibility $\chi_{\rm ac}(T)$ is measured in several runs up to $\sim4$\,GPa. 
The temperature sweeps for the 1st run are shown in Fig.\,S3. Clear diamagnetic signals due to superconducting transitions are observed below the pressure-dependent $T_{\rm c}$. We note that the magnitude of diamagnetic signals at low temperature does not show strong  pressure dependence, suggesting that bulk superconducting property persists up to $\sim3$\,GPa, where $T_{\rm c}$ is enhanced significantly from the ambient pressure value.

\begin{figure}[t]
	\includegraphics[width=0.7\linewidth]{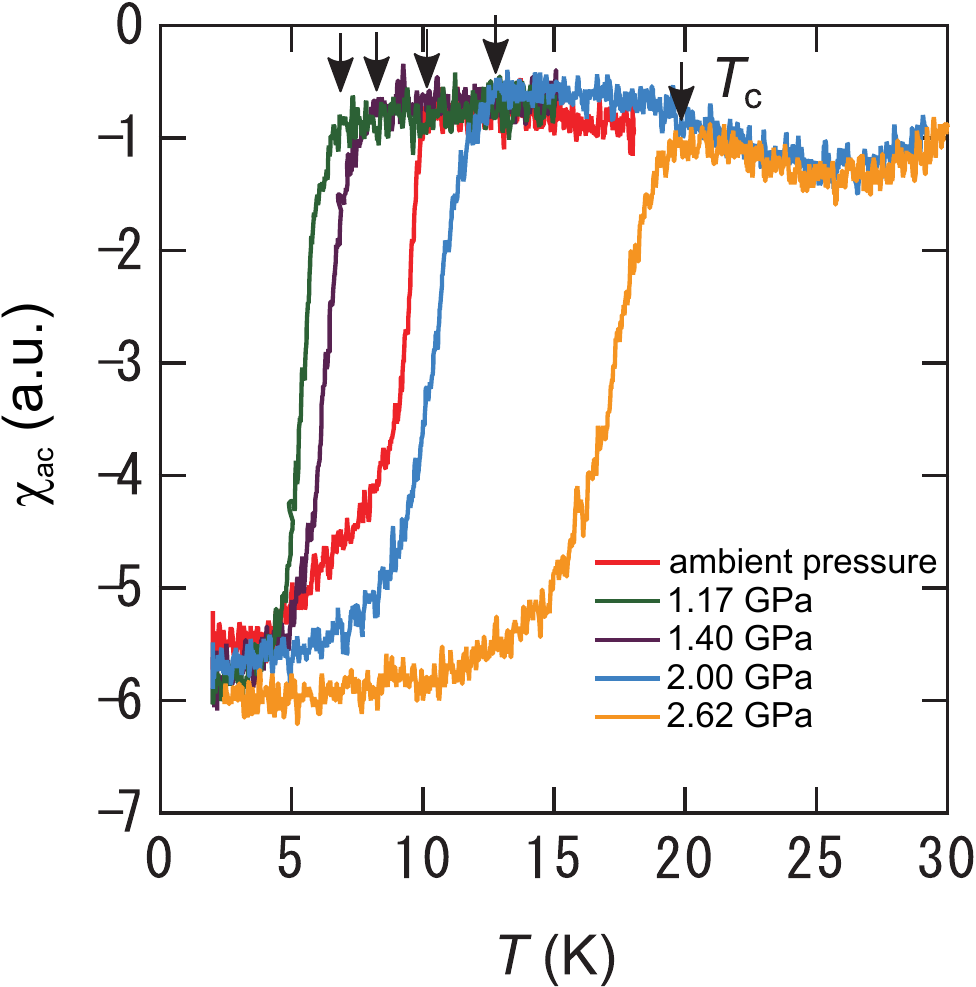}
	\caption{
		\textbf{Temperature dependence of ac susceptibility under pressure in FeSe$\boldsymbol{_{1-x}}$S$\boldsymbol{_{x}}$ for $\boldsymbol{x=0.12}$.} The measurements are performed in the increasing order of pressure. The arrows indicate $T_{\rm c}$, which is determined by the onset of the diamagnetic signal.
	}
\end{figure}

\begin{figure}[b]
	\includegraphics[width=0.7\linewidth]{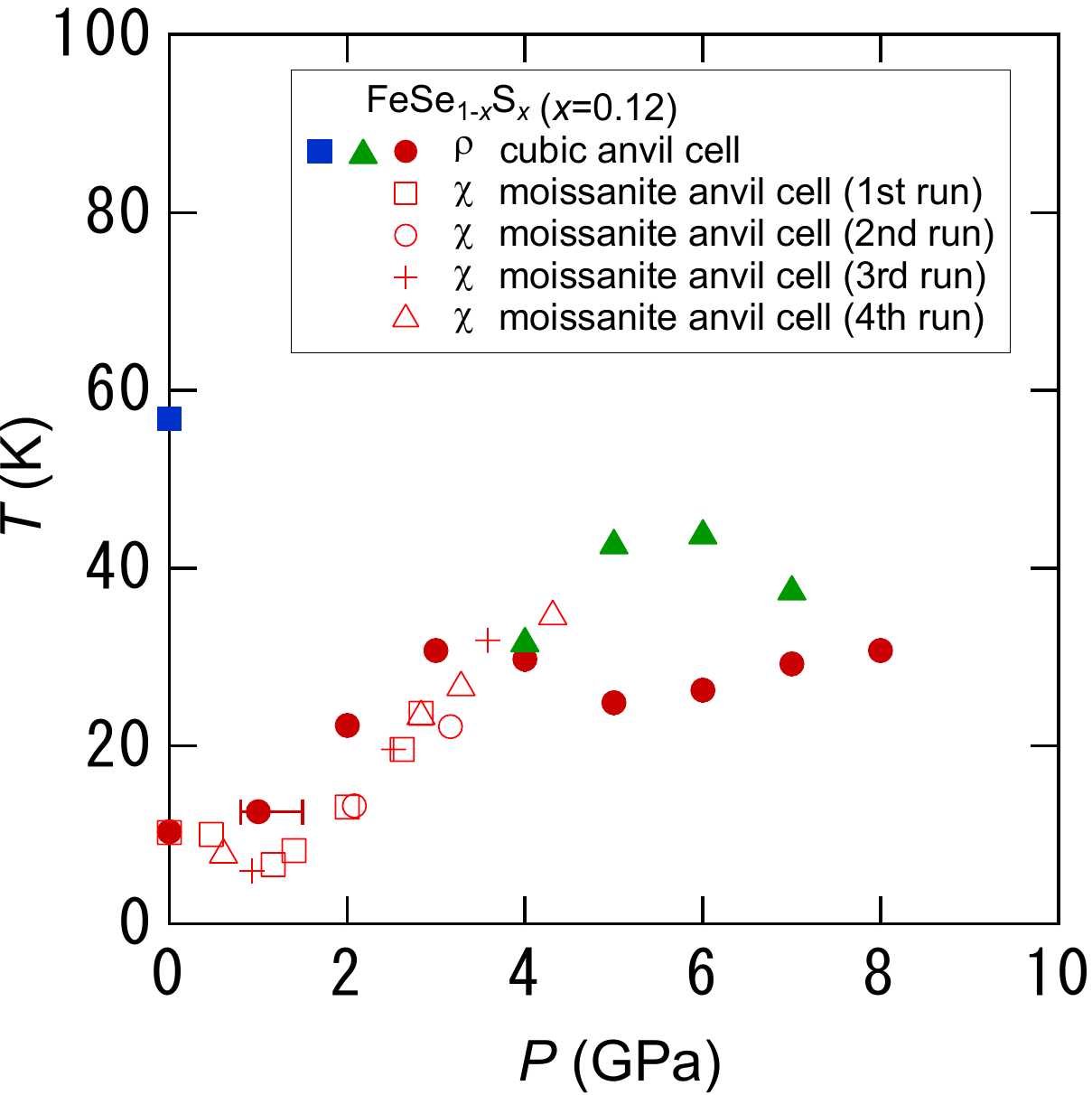}
	\caption{
		\textbf{Pressure dependence of $\boldsymbol{T_{\rm c}}$ determined by the ac susceptibility compared with the resistivity-determined phase diagram of $\boldsymbol{x = 0.12}$.}
		$T_{\rm c}$ determined by the ac susceptibility measurements for several different runs (open symbols) are plotted as a function of pressure estimated at room temperature.  
		For comparison the nematic (blue squares), magnetic (green triangles) and superconducting transition temperatures (red closed squares) are also shown (from Fig.\,1). The likely source of the difference between the pressure values from two different cells is discussed in the text.
	}
\end{figure}

\begin{table*}[t]
	\begin{center}
		\caption{{\bf Lattice parameters of FeSe$\boldsymbol{_{1-x}}$S$\boldsymbol{_{x}}$.}
			Lattice constants $a$, $c$, and the chalcogen height $h_{\rm Ch}$ as a function of S-content $x$ at ambient pressure (left), together with the data under pressure for FeSe polycrystals taken from Millican {\it et al.} \cite{Millican09} (right).
		}
		\label{tab1}
		\vspace{2mm}
		\scalebox{1.15}{
			\begin{tabular}{|c|c|c|c|c||c|c|c|c|c|}
	\hline
	$x$ &  0 & 0.044 & 0.073 & 0.136 & $P$ (GPa) & 0 & 0.2 & 0.4 & 0.6 \\
	\hline
	$a$\,(\AA) & 3.7742 & 3.7614 & 3.7667 & 3.7624 & $a$\,(\AA) & 3.7658 & 3.7610 & 3.7555 & 3.7501 \\
	\hline
	$c$\,(\AA) & 5.5279 & 5.5021 & 5.4913 & 5.4675 &$c$\,(\AA) &  5.4988 & 5.4794 & 5.4598 & 5.4398 \\
	\hline
	$h_{\rm Ch}$\,(\AA) & 1.4754 & 1.4702 & 1.4670 & 1.4603 & $h_{\rm Ch}$\,(\AA) &  1.4643 & 1.4668 & 1.4676 & 1.4682 \\
	\hline
	
\end{tabular}				}
	\end{center}
\end{table*}

The pressure dependence of $T_{\rm c}$ determined by the ac susceptibility method is compared with the resistivity-determined phase diagram for $x=0.12$ in Fig.\,S4. The pressure dependence of the diamagnetic $T_{\rm c}$ for several runs shows a similar trend with that of the zero-resistivity temperature. We note that the pressure for the ac susceptibility measurements is estimated by ruby fluorescence spectrum at room temperature, and thus it is likely that at low temperatures the actual pressure is lower than this estimate. Furthermore, at low pressures it is trickier to control the pressure of the cubic anvil cell. 
Considering the difference of pressure techniques between self-clamped type moissanite anvil cell for ac susceptibility measurements and the constant-loading type cubic anvil cell for resistivity measurements, the $T_{\rm c}(P)$ results from these two different measurements are in fairly good agreement. Most importantly, the observed trend of $T_{\rm c}(P)$ from two techniques is genuine. From these results we conclude that a new, bulk high-$T_{\rm c}$ superconducting phase exists near the lower boundary of the pressure-induced magnetic dome.

Owing to the capability of finer tuning of pressure for the clamp-type cell, we find a minimum at $P\sim1$\,GPa in the susceptibility-determined $T_{\rm c}(P)$ curve. This nonmonotonic $P$-dependence of $T_{\rm c}$ resembles that for pure FeSe in the low-pressure range, where the nematic transition is fading. This suggests that the $T_{\rm c}$ minimum found here for $x = 0.12$ is associated with the fate of the nonmagnetic nematic transition, consistent with our interpretation.

\subsection{Lattice parameters}

For $x=0.08$, the Bragg peaks measured by synchrotron X-ray diffraction show contrasting behaviours between 3.0\,GPa where high-$T_{\rm c}$ superconductivity appears and 4.9\,GPa where pressure-induced magnetism is evidenced (Fig.\,3). In Fig.\,S5, we compare quantitatively the temperature dependence of $(220)$ Bragg peak width between these two pressures. At high temperatures, both data show similar peak width with no splitting, but at low temperature the 4.9-GPa data shows a clear splitting. The orthorhombicity rapidly develops below $T_{\rm s}\approx 41$\,K, implying the first-order nature of the tetragonal-to-orthorhombic transition, which is similar to the case for pressure-induced SDW phase of FeSe \cite{Kothapalli16}. Except near the transition, the widths of the split peaks in the orthorhombic phase (triangles in Fig.\,S5b) are both comparable to that of the tetragonal phase. Comparing with the results of FeSe by Kothapalli {\it et al.} \cite{Kothapalli16}, the orthorhombicity $\delta\approx 1.4\times 10^{-3}$ at low temperatures is about half of that at ambient pressure ($\delta\approx 2.7\times 10^{-3}$) but is close to that at 3.1\,GPa ($\delta\approx 1.7\times 10^{-3}$). 

The comparisons of lattice parameters between the $x$-dependence at ambient pressure and the $P$-dependence for $x=0$ are listed in Table\,SI (see also Fig.\,4). The $P$-dependence data are taken from results for FeSe polycrystals by Millican {\it et al.} \cite{Millican09}. The slight differences in the parameters for FeSe at ambient pressure between our single crystals and their polycrystals are possibly due to small deficiency of Se atoms in their samples.

\begin{figure}[b]
	\includegraphics[width=\linewidth]{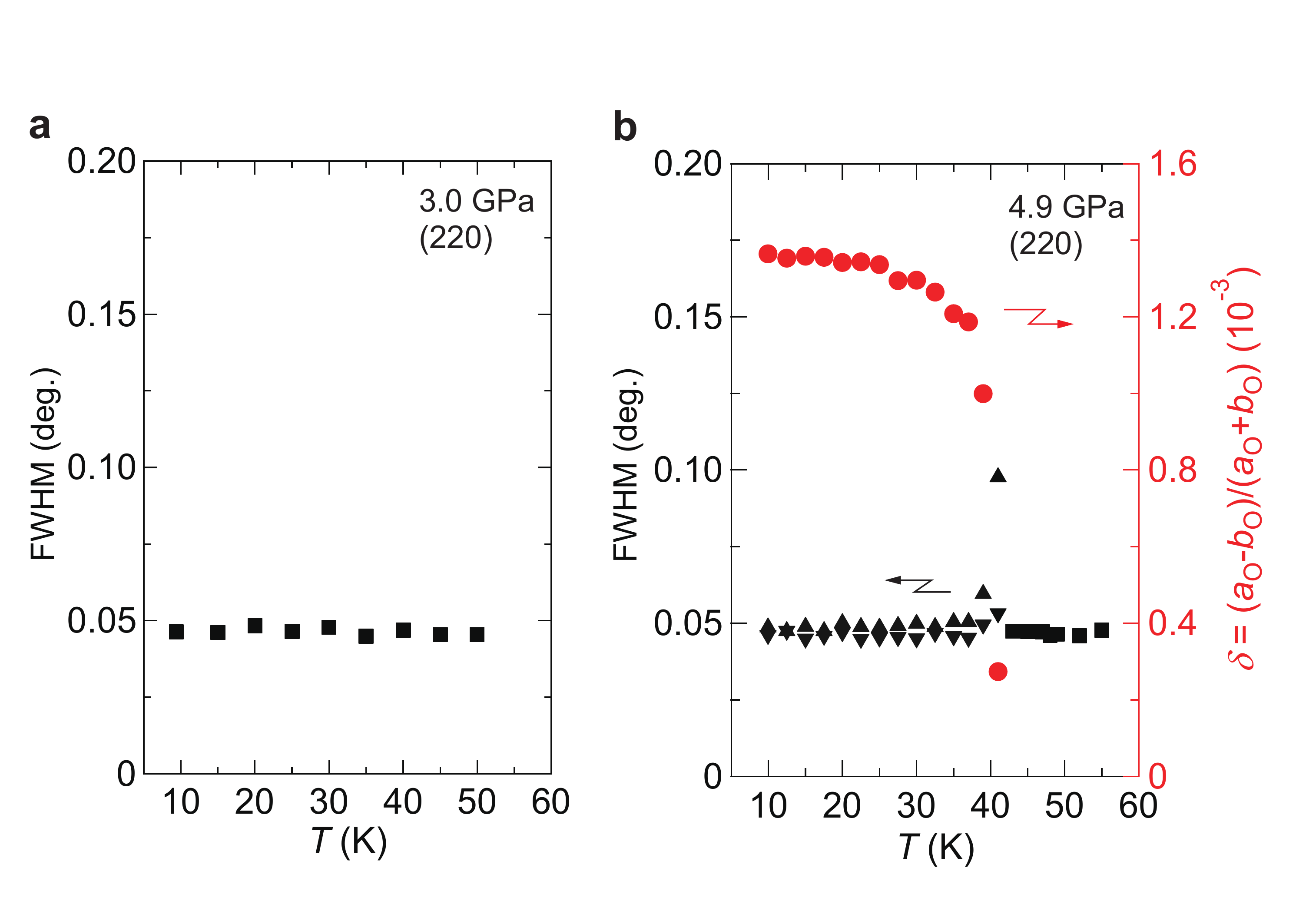}
	\caption{
		\textbf{Temperature dependence of Bragg-peak width and orthorhombicity in FeSe$\boldsymbol{_{1-x}}$S$\boldsymbol{_{x}}$ for $\boldsymbol{x=0.08}$ at high pressure.} {\bf a}, Full width at half maximum (FWHM) of the $(220)$ Bragg peak as a function of temperature at 3.0\,GPa. {\bf b}, FWHM in the high-temperature tetragonal phase (black squares and triangles, left axis) and the orthorhombicity $\delta=(a_{\rm O}-b_{\rm O})/(a_{\rm O}+b_{\rm O})$ below $T_{\rm s}\approx 41$\,K determined by the splitting of the $(220)$ Bragg peak (red circles, right axis) are plotted as a function of temperature for 4.9\,GPa.
	}
\end{figure}

\end{document}